\documentclass[a4paper,12pt]{article}
\usepackage[utf8]{inputenc}
\usepackage{amsmath}
\usepackage{amssymb}
\usepackage{mathrsfs}
\usepackage{chemarrow}
\usepackage{times}
\usepackage{graphicx}
\begin{document}

\def\mD{{\bf D}}
\def\rd{{\rm d}}
\def\mF{{\bf F}}
\def\mQ{{\bf Q}}
\def\vJ{{\bf J}}
\def\rvB{{\bf B}}
\def\bplus{{\bf +}}
\def\bplus{\mbox{\boldmath$+$}}
\def\pa{{\bf a}}
\def\pp{{\bf p}}
\def\vx{{\bf x}}
\def\vX{{\bf X}}
\def\vz{{\bf z}}
\def\vnu{\mbox{\boldmath$\nu$}}


\title{Nonlinear Stochastic Dynamics of Complex Systems, I:
A Chemical Reaction Kinetic Perspective with
Mesoscopic Nonequilibrium Thermodynamics
}

\author{Hong Qian\footnote{hqian@u.washington.edu}\\[6pt]
Department of Applied Mathematics\\
University of Washington, Seattle\\
WA 98195-3925, U.S.A
}

\maketitle

\begin{abstract}
We distinguish a mechanical representation of the world in terms of point masses with positions and momenta and the chemical representation of the world in terms of populations of different individuals, each with intrinsic stochasticity, but population wise with statistical rate laws in their syntheses, degradations, spatial diffusion, individual
state transitions, and interactions. Such a formal kinetic system in a small volume $V$, like a single cell, can be rigorously treated in terms of a Markov process describing its nonlinear kinetics as well as nonequilibrium thermodynamics at a mesoscopic scale. We introduce notions such as open, driven chemical systems, entropy production, free energy dissipation, etc. Then in the macroscopic limit, we illustrate how two new ``laws'', in terms of a generalized free energy of the mesoscopic stochastic dynamics, emerge. Detailed balance and complex balance are two special classes of ``simple'' nonlinear kinetics. Phase transition is intrinsically related to multi-stability and saddle-node bifurcation phenomenon, in the limits of
time $t\rightarrow\infty$ and system's size $V\rightarrow\infty$. Using this approach, we re-articulate the notion of inanimate equilibrium branch of a system and nonequilibrium state of a living matter, as originally proposed by Nicolis and Prigogine, and seek a logic consistency between this viewpoint and that of P. W. Anderson and J. J. Hopfield's in which macroscopic law emerges through symmetry breaking.
\end{abstract}

\tableofcontents

\section{Introduction}

	This is the Part I of an intended series on a comprehensive theory of complex systems \cite{part-II}. 
Taking living biological cells as an archetype and biochemical kinetic approach as a paradigm, we consider a system {\em complex} if it contains many interacting
subpopulations of individuals that undergo non-deterministic state transitions, and 
it interacts with its environment through active transports of matters and energy, or information.  For examples, tumor is a community of heterogeneous individual 
cells; economy concerns with individual agents, and ecology deals with
various biological organisms.  

	The above ``definition'' of a complex system immediately reminds us 
several major areas of studies: nonequilibrium thermodynamics, statistical physics, 
nonlinear dynamics, stochastic processes, and information theory are some of
the widely acknowledged.  Ours, however, is not a ``new'' theory, 
rather it is a coherent narrative that synthesizes a wide range of methodologies
and thoughts mentioned above.  One can easily find in our
writting significant influences from 
the various theories on complexity: the Brussels school of 
thermodynamics that describes nonequilibrium phenomena in terms of chemical
affinity and entropy production 
\cite{deDonder,np-book,nicolis-book}; the theory of synergetics which articulates
the ideas of nonequilibrium potential and phase transition, as well as slaving
principle as an emergent phenomenon near a critical point \cite{haken_83,haken_00}; 
the notion of symmetry breaking from the phase transition lore in condensed matter physics \cite{pwanderson,jjh,fw_94,laughlin_wolynes};
catastrophe theory in connection to nonlinear bifurcation
\cite{thom-book,zeeman-88}.

	The unique features of our approach are as follows:  First, we have
two concrete examples of complex chemical systems in mind: A single protein
molecule, which consists of a large collection of heterogeneous atoms, 
in an aqueous solution as a non-driven (closed) chemical system \cite{qian-ps}, 
and a single cell, which consists of a large number of different biomolecular
species, in a culture medium as a driven open chemical system \cite{qian-arbp}.
While closed and open chemical systems have fundamentally different 
thermodynamics, some of the key aspects of stochastic nonlinear dynamics 
are remarkably consistent:  1) Nonlinearity gives rise to discrete states, 
e.g., conformational states of a protein and epi-genetic states of a cell; 2)
stochasticity yields discrete transitions among the states on a much longer
time scale. Both systems can be mathematically described in terms of
stochastic nonlinear kinetics.

	The second feature of our approach is following what the stochastic, nonlinear mathematics
tells us \cite{guicciardini-book}.  For examples: i) For many of the stochastic models of open chemical
system, one can prove the existence of a unique, ergodic steady state probability
distribution, even though its computation is often challenging toward which  
significant past research has been directed.  But the
latter should not prevent one to develop a theory based on such a 
nonequilibrium steady state (NESS) potential function.  Indeed, we can show
that the mere recognition of its existence provides great logical consistency
and theoretical insights in a theory.   ii) For Markov stochastic processes under 
a set of rather weak conditions, one can show a set of mathematical 
theorems that have remarkable resemblance to the 
theory of chemical thermodynamics, {\it \`{a} la} Gibbs, and
Lewis and Randall \cite{lewis-randall-book,guggenheim-book}.  
iii) For any stochastic dynamics with a 
macroscopic bistability as fluctuations tending to zero ($\varepsilon\rightarrow 0$), 
symmetry breaking in the limit of $\varepsilon\rightarrow 0$ followed by time $t\rightarrow\infty$ is a necessary consequence of the catastrophe
in the nonlinear dynamics, obtained from $\varepsilon\rightarrow 0$ precedes
$t\rightarrow\infty$.  While these mathematical results stand on their own, 
their significance to our understanding of complex systems requires a 
narrative that is constructed based on all the past theories, thoughts, 
and discussions.

	The stochastic nonlinear dynamic description of complex systems 
embodies two of the essential ingredients of Darwin's theory of 
biological evolution: chance, variation, diversity on the one hand and 
necessity, selection on the other \cite{monod-book,haken_83,ao-05}.
The chemical kinetic theory of living cells based on the  
Delbr\"{u}ck-Gillespie process \cite{leontovich-35,delbruck,kurtz-72,gillespie,erdi-book}
allows one to see how various abstract ``forces of nature'' emerge and
play out in a complex system, with individual players as simple, or as
complex, as macromolecules.

\subsection{Mechanics and chemistry}

	Mechanics and chemistry offer two very different 
perspectives of a complex system: the former considers the world
made of featureless individuals, the {\em point masses},\footnote{A more profound insight is that no matter how complex interactions among a collection of point masses are, its center of mass behaves 
as a single featureless point mass.} with precise positions and velocities; the latter, however, entertains a world made of many types of individuals, each with different internal  characteristics.  
Due to the uniqueness of space and time, each and every point mass
is unique.  On the other hand, modern studies have shown that 
individual single molecules in aqueous solution behavior 
differently: They have {\em individualities} \cite{xie98,wemoerner}.
More interestingly, when dealing with many-body systems, even
fluid mechanics and quantum mechanics  turn their representations
from tracking the state of individual particles to 
counting the numbers: the switching from Lagrangian to Eulerian descriptions in the former and second quantization in the latter.

	The ``chemical kinetic view'' of the complex world goes much beyond 
chemistry: Reaction $S+I\rightarrow 2I$ is known as autocatalysis
in chemistry, with the $S$ being a reactant and $I$ being its product
which also serves as a catalyst for a transformation of an $S$ to an $I$.
The same ``kinetic scheme'' also describes infection in an 
epidemiological dynamics, with $S$ and $I$ representing susceptible
and infectious individuals, respectively.  In fact, we argue that
much of the complex systems beyond chemistry contain interacting 
subpopulations of individuals that undergo non-deterministic state
transitions.  The chemical view actually offers a reasonable organization
of P. W. Anderson's $X$ and $Y$ in the hierarchical structure of
science \cite{pwanderson}:
\[
    \begin{array}{ccc}
             X  &&  Y  \\
    \text{many-body physics}  &&   \text{particle physics}  \\
     \text{chemistry}  &&   \text{many-body physics}  \\
     \text{molecular biology}  &&   \text{chemistry}  \\
     \text{cell biology}  &&   \text{molecular biology}  \\
          \vdots &&  \vdots \\
	     \text{psychology}  &&  \text{physiology}  \\
     \text{social sciences}  &&   \text{psychology} 
	\end{array}
\]
We note that each $X$ in the left column consists of 
collections of ``individuals'' characterized by the $Y$ on 
the right.

\section{Chemical kinetics and chemical thermodynamics}
\label{sec:}

\subsection{Chemical reaction as an emergent phenomenon}

	Is a biochemical macromolecule, a protein, immersed in an aqueous 
solution simple or complex?  The answer to this question depends on
one's perspective, and time scale \cite{fw_94}.  From the standpoint of molecular 
physics and in terms of the large number of atoms that are constantly 
in collisions with each other and with solvent molecules, 
this is a quite complex system which has its own 
emergent phenomenon.

	The mechanical motions described by a Newtonian molecular 
dynamics (MD) with solvent molecules being explicitly treated, usually on the time scale of
femtosecond (10$^{-15}$ sec.), exhibit great complexity.  Yet, the 
notions of ``reaction coordinate'' (e.g., order parameter) and 
``transition state'' (e.g., critical state), plus a single rate constant
which is on the microsecond to millisecond time scale,
fully describe the exponential law for an elementary, unimolecular
chemical reaction.  This emergent phenomenon has been
mathematically described in H. A. Kramers' stochastic, nonlinear theory of 
energy barrier crossing \cite{kramers}, and the rigorous mathematics of 
Freidlin-Wentzell theory \cite{fw-book,anton-bovier,nils-berglund}.

\subsection{Chemical kinetics and elementary reactions}

	Just as point mass is a fundamental concept in mechanics,
{\em elemetary reaction} is a fundamental concept in chemical 
kinetics: A reaction like
\[
       \nu^+_1 A_1 + \nu^+_2 A_2 + \cdots
       \nu^+_N A_n \longrightarrow \nu^-_1A_1 + \nu^-_2 A_2 + \cdots
       \nu^-_N A_n
\]
is said to be elementary if the discrete event of transformation
has an exponentially distributed waiting time with rate $r$,
which is a function of numbers of the reactants, 
$n_1,n_2,\cdots,n_N$, in a reaction vessel with volume $V$.
If the rate $r$ has the form of
\begin{equation}
     r = kV\prod_{\ell=1}^N \left(\frac{n_{\ell}
           (n_{\ell}-1)\cdots (n_{\ell}-\nu_{\ell}^++1)}{V^{\nu_{\ell}^+}}\right),
\label{meso-flux}
\end{equation}
in which $n_\ell$ is the number of molecule $A_\ell$ in the vessel.
then we said the reaction follows the law of mass action with rate constant $k$.  
In a macroscopic sized system, in terms of concentrations
$x_1,x_2,\cdots,x_N$, the reaction flux in (\ref{meso-flux}) 
becomes
\begin{equation}
     J = \frac{r}{V} = k\prod_{\ell}^N x_{\ell}^{\nu^+_{\ell}},
\end{equation}
in which $x_\ell=n_{\ell}/V$ is the concentration of the reactant
$A_\ell$.   For example, the nonlinear kinetic system
\begin{equation}
     S + I \overset{k_1}{\longrightarrow} 2I,  \  \
       I \overset{k_2}{\longrightarrow} R,
\end{equation}
has a macroscopic kinetic equation following the law of 
mass action
\begin{equation}
   \frac{\rd x}{\rd t} = k_1xy-k_2x,  \  \
    \frac{\rd y}{\rd t} = -k_1xy,  \ \
     \frac{\rd z}{\rd t} = k_2x,
\end{equation}
where $x$, $y$, and $z$ are the concentrations of chemical
species $I$, $S$, and $R$, respectively.  In the mathematical
theory of infectious diseases, however, 
the same set of equations represent the SIR model
\cite{jdmurray}.  The population dynamics of species in 
an infectious disease, in ecological dynamics, and 
chemical species in a rapid stirred reaction vessel, 
actually share a great deal of commonality: They are 
all complex nonlinear systems consisting of heterogeneous
interacting individuals.

\subsection{Mesoscopic description of chemical kinetics}
\label{sec:meso-d-ck}

	The readers are referred to the earlier review articles 
\cite{qian-bishop-ijms,liang-qian-jcst} and the 
recent text \cite{erdi-book}.

\section{Nonequilibrium thermodynamics (NET)}
\label{sec:net}

	The notion of elementary reactions as presented 
above is fundamentally a stochastic one in which
a chemical reaction in an aqueous solution is represented by a 
rare event in the many-body, heterogeneous atomic system.
Each reaction as a single random event is the focus of the Kramers'
theory whose prediction is that, when the barrier 
is high, all the complex atomic motions in a unimolecular 
transition can be represented by
a single parameter, the rate constant $k$.  For bimolecular
reactions, a diffusion based theory in three-dimensional
physical space was first 
proposed by M. von Smolochowski (1917), and further developed
by Collins and Kimball (1949), A. Szabo and his coworkers 
\cite{sss-80,agmon-szabo-90}. 

	While Kramers' theory connects stochastic 
molecular state transitions to the motions of the constitutive atoms, 
a cell consists of a great many macromolecules and 
biochemical reactions.  The Delbr\"{u}ck-Gillespie process describes 
the stochastic kinetics of such a system, with two complementary
representations: Either in terms of the random, fluctuating number 
of all molecular species, $\vec{n}=$
$\big[n_1(t),n_2(t),\cdots,n_N(t)\big]$ 
that follows Gillespie algorithm \cite{gillespie}, or in terms of 
the probability distribution
$p(m_1,\cdots,m_N;t)=$ $\Pr\{n_1(t)=m_1,\cdots,n_N(t)=m_N\}$ 
that satisfies the Chemical Master Equation (CME) \cite{delbruck,kurtz-72}.
This theory of mesoscopic stochastic chemical kinetics has rapidly 
become a mature subject in recent years \cite{erdi-book}.

	We use the term ``mesoscopic'' in partial agreement with
van Kampen \cite{vankampen-book}, Ch. III, p. 57,  who stated 
that ``[t]he stochastic description in terms of 
macroscopic variables ... It comprises both the deterministic
laws and the fluctuations about them''.  There is an important
difference, however: What implicitly assumed by van Kampen
and physicists of his time was  a top-down, macroscopic law known 
first followed by fluctuations theory; it is a phenomenological
approach pioneered by Einstein.  

The stochastic chemical kinetic 
theory is different.  With a foundation laid by Kramers' theory, 
a stochastic chemical kinetic model is {\em mechanistic}.
The deterministic laws then can be mathematically derived, as
has been shown by T. G. Kurtz \cite{kurtz-72}.   This 
conceptual distinction also leads to an essential practical
difference: The ``noise structure'' in the traditional top-down
models can only be determined by additional assumptions, 
e.g., fluctuation-dissipation relation for equilibrium fluctuations.
The noise structure in a  Delbr\"{u}ck-Gillespie model is completely
specified by the chemical kinetics. 

	The most general setup for a mesoscopic chemical kinetics
in a rapidly stirred reaction vessel with volume $V$ considers
$N$ chemical species and $M$ elementary chemical reactions: 
\begin{equation}
     \nu^+_{\ell 1}X_1+\nu^+_{\ell 2}X_2 + \cdots
     \nu^+_{\ell N}X_N  \  \  \underset{r_{-\ell}}{\overset{r_{+\ell}}{\rightleftharpoonsfill{26pt}}}   \  \
      \nu^-_{\ell 1}X_1+\nu^-_{\ell 2}X_2 + \cdots
     \nu^-_{\ell N}X_N,
\label{rxn}
\end{equation}
in which both $r_{\pm\ell}$ are functions of $\vec{n}$
and $V$, $1\le \ell\le M$. $\nu^{\pm}_{\ell j}$ are stoichiometric
coefficients for the forward and the backward reactions $\pm\ell$.

In applied mathematics, compared with what we know abut 
ordinary differential equations beyond the existence and uniqueness 
theorems, we currently know very little about the Delbr\"{u}ck-Gillespie process.

\subsection{Gibbs' chemical thermodynamics}

	It is widely agreed upon that thermodynamic behavior is an 
emergent phenomenon of systems with a large number of 
components.  L. Boltzmann's attempt to provide macroscopic
thermodynamics with a Newtonian mechanical foundation has 
provided continuous inspiration for understand complexity.  
Consider a mechanical system, a box of gas with volume $V$
and number of particles $N$, described by Hamiltonian dynamic 
equation $\dot{p}=\partial H(p,q/\partial q$ and
$\dot{q}=-\partial H(p,q)/\partial p$.
According to Boltzmann's fundamental insight \cite{gallavotti-book},
a macroscopic thermodynamic state is a state of motion, the
entire level set of $H(p,q;V,N)=E$, which is determined by the initial
condition of the differential equation.  Then 
entropy $S=k_B\ln\Omega(E,V,N)$ where $\Omega(E,V,N)$ is the
phase volume contained by the level set.
Thus there exists a definitive function 
$S=S(E,V,N)$.  It then follows from elementary calculus:
\begin{equation}
  \rd E = \left(\frac{\partial E}{\partial S}\right)_{V,N}\rd S
             +\left(\frac{\partial E}{\partial V}\right)_{S,N}\rd V
             +\left(\frac{\partial E}{\partial N}\right)_{S,V}\rd N
\label{gibbs-eq}
\end{equation}
in which $(\partial E/\partial S)_{V,N}=T$ is identified as
temperature, $(\partial E/\partial V)_{S,N}=-p$ is
identified as pressure, and $p\rd V$ is mechanical work.
$(\partial E/\partial N)_{S,V}=\mu$ is called chemical 
potential.

Consequently thermodynamic quantities such as $T$, $p$, and $\mu$ 
are emergent concepts themselves.  However, for both $T$ and $p$
clear {\em mechanical interpretation} have been found: 
$T$ being mean kinetic energy and $p$ the momentum 
transfer upon collision of gaseous molecules on the wall of the box.
However, one yet to find a clear mechanical interpretation for
$\mu$.  

The chemical potential $\mu$ has a $\mu^o$ part and a 
log concentration (or mole fraction) part.  The current molecular theory of $\mu^o$ is based 
on molecular mechanics, which treats atoms in a molecule as point mass etc.
In the past, molecules has been understood through mechanics, classical or 
quantum.

The entire living phenomena are mostly 
driven by $\Delta\mu$, not $\Delta T$ or $\Delta p$.
Chemistry kinetics offers a concrete example of complex 
systems.

\subsection{The source of complexity}

	Biological systems are archetypes of complex system.
Why biological and complex systems look so different from 
those in physics? J. J. Hopfield \cite{jjh}, together with many other
condensed matter physicists
\cite{haken_00,nicolis-book,bialek-book}, have all 
pointed to the information content of a system, and the 
notion of {\em symmetry breaking} as the key elements of
complex systems \cite{pwanderson,fw_94,laughlin_wolynes}.
The notion of symmetry breaking, in a
broad sense, is best illustrated in a mesoscopic kinetics 
with finite volume $V$: When $V$ is small, the Markov 
dynamics is ergodic.  Multi-stability in this system is 
represented by multiple peaks of the stationary probability
distribution $p^{ss}(n_1,n_2,\cdots,n_N;V)$ 
\cite{np-book,haken_83,zeeman-88}.  
However, 
when $V\rightarrow\infty$, the different peak regions correspond
to different basins of attraction according to the systems of
ordinary differential equations (ODEs) that follow
the law of mass action.   The dynamics of the
deterministic ODEs has broken ergodicity: Dynamics starts in one
basin will never go to another basin.  More interestingly, if one takes
the limit of $V\rightarrow\infty$ after $t\rightarrow\infty$ in the
mesoscopic kinetic model, it predicts the system has only a 
single attractor with probability 1.   In fact, the stationary probability
distribution
\begin{subequations}
\begin{eqnarray}
    V^{-1}p^{ss}(x_1V,x_2V,\cdots,x_NV;V)
     &\approx& \Xi^{-1}(V)e^{-V\varphi(x_1,x_2,\cdots,x_N)},
\\
                 \Xi(V) &=& \int_{\mathbb{R}^N}
                e^{-V\varphi(\vec{x})}\rd\vec{x},
\end{eqnarray}
\end{subequations}
in which $\varphi(\vec{x})$ has a global minimum of 0, at which
the entire probability will concentrate when 
$V\rightarrow\infty$ \cite{aqtw}.  Phase transition occurs when
the global minimum of $\varphi(\vec{x})$ is not unique.

\subsection{Macroscopic NET of continuous media}

A summary of the standard formalism of macroscopic 
nonequilibrium thermodynamics (NET) in continuous medium, as
introduced in \cite{dm-book}, can also be found in \cite{qkkb-16}.
In this theory, the existence of a special macroscopic function 
$s(x,t)=\Theta\big[u(x,t),\{c_i(x,t)\}\big]$, the 
instantaneous entropy density, is hypothesized via 
the {\em local equilibrium assumption}, which yields 
\begin{equation}
             \frac{\partial s(x,t)}{\partial t} 
      = T^{-1}(x,t)\left[\frac{\partial u(x,t)}{\partial t} 
-\sum_{i=1}^K \mu_i(x,t)\frac{\partial c_i(x,t)}{\partial t} \right],
\label{lea}
\end{equation}
in which $u(x,t)$ is internal energy density, $c_i(x,t)$ is the
concentration of  the $i$th chemical species, $T$ and $\mu_i$ 
are temperature and chemical potentials.  The theory of NET then
proceeds as follows:

($ii$) Establishing continuity equations for $u(x,t)$ and $c_i(x,t)$:
$\partial u_t = -\partial_xJ_u(x,t)$, 
$\partial (c_i)_t = -\partial_x J_i(x,t) + \sum_{j=1}^M \nu_{ji}r_j$,
where $J_u$ and $J_i$ are energy and particle fluxes in space,
and $r_j$ is the rate of the $j$th reaction with stoichiometric
coefficients $\nu_{ji}$;

($iii$) Substituting the $J$'s and $r$'s into (\ref{lea});

($iv$) Grouping appropriate terms to obtain the 
density of entropy production rate, $\sigma(x,t)$,
as  ``transport flux $\times$ thermodynamics force'', 
{\it \`{a} la} Onsager.  The remaining part is 
the entropy exchange flux $J_s(x,t)$:
$\partial s(x,t)/\partial t = \sigma(x,t)  -\partial_x J_s(x,t)$.

\subsection{Mesoscopic NET}

	The $x$ in the above macroscopic NET represents the real three-dimensional 
physical space.  One can apply a similar approach to a dynamics 
in phase space, as Bergmann and Lebowitz's stochastic Liouville dynamics
\cite{BL-55,LB-57}, or the space of internal degrees of 
freedom initiated by Prigogine and Mazur 
\cite{prigogine-mazur-53,mazur-98,mazur-99,rubi-jpc-05}.
In these cases, the theory of mesoscopic NET proceeds 
as follows:  

i)  Introducing the entropy, or free energy, as a functional of
the probability distribution $p_{\alpha}(t)$, $\alpha\in\mathscr{S}$
for a discrete system, or probability density function $f(\vx,t)$;
$\vx\in\mathbb{R}^n$, for a continuous system.  The introduction of 
entropy function in the mesoscopic theory does not rely on the local 
equilibrium assumption.  Rather, one follows the fundamental idea 
of L. Boltzmann: The entropy is a functional characterizing the 
{\bf\em statistics} of phase space.  One of such functionals 
for stochastc processes is the Shannon entropy.

ii) Establishing the continuity equation in phase space
connecting the probability distribution and probability flux:
\begin{subequations}
\begin{eqnarray}
	\frac{\rd p_{\alpha}(t)}{\rd t} &=& 
        \sum_{\beta\in\mathscr{S}} \Big( J_{\beta\rightarrow\alpha}(t)
          - J_{\alpha\rightarrow\beta}(t) \Big),
\\
    \frac{\partial f(\vx,t)}{\partial t} &=& -\nabla_{\vx}\cdot
               \vJ(\vx,t),
\end{eqnarray}
\end{subequations}

iii) Computing the time derivative of $\tfrac{\rd}{\rd t}S[p_n(t)]$, 
or $\tfrac{\rd}{\rd t}S[f(\vx,t)]$, following the chain rule for differentiation;

iv) Setting up entropy production rate $\sigma$ in terms 
of bi-linear products of  ``thermodynamic fluxes'' and ``thermodynamic
forces''.

\subsection{Mesoscopic NET as a foundation of stochastic
dynamics}
\label{sec:mesonet}

	While the mathematics in the theory of mesoscopic 
nonequilibrium thermodynamics (meso-NET), pioneered by Prigogine 
and Mazur \cite{prigogine-mazur-53,mazur-98}, and fully 
developed by Rub\'{i} and coworkers \cite{rubi-jpc-05}, are 
essentially the same as in the theory of stochastic Liouville dynamics 
(Sec. \ref{sec:sld}) and the recently developed stochastic 
thermodynamics (Sec. \ref{sec:meso-stoch-net}), the scientific 
narratives are quite different.  The goal for the former is to use meso-NET 
to {\em derive} the dynamics equations for fluctuations of macroscopic
quantities \cite{mazur-99}; more specifically, Fokker-Planck equation for probability
density function $f(\vx,t)$.  To illustrate this, one starts with the $f(\vx,t)$
that necessarily satisfies a continuity equation in a phase space:
\begin{subequations}
\begin{equation}
     \frac{\partial f(\vx,t)}{\partial t} = 
                 - \sum_{i}\frac{\partial}{\partial x_i} J_i(\vx,t),
\label{conti-eq}
\end{equation} 
and non-adiabatic entropy production, or free energy 
dissipation \cite{ge-qian-10}:
\begin{equation}
  -\frac{\rd}{\rd t} \int_{\mathbb{R}^n} f(\vx,t)\ln\left(\frac{f(\vx,t)}{
        \pi(\vx)}\right)\rd\vx = \int_{\mathbb{R}^n} 
          f(\vx,t)\sigma^{(na)}(\vx,t)\rd\vx, 
\end{equation}
in which the {\em local density} of non-adiabatic entropy production rate
\begin{equation}
  \sigma^{(na)}(\vx,t) = -\sum_{i} J_i(\vx,t)\frac{\partial}{\partial x_i}
              \ln\left(\frac{f(\vx,t)}{\pi(\vx)}\right),
\end{equation}
where $\vJ(x,t)$ is the flux and the 
$\nabla\ln\big[f(x,t)/\pi(x)\big]$
is the thermodynamic force: the gradient of the local
chemical potential function.  One also notices that the thermodynamic
force $\nabla\ln\big[f(\vx,t)/\pi(\vx)\big]$ can be further decomposed 
into $\vX+\nabla\ln f(\vx,t)$ where $\vX=-\nabla\ln\pi(\vx)$ is
the thermodynamic variables conjugate to $\vx$.  Readers with a 
college chemistry background should recognize this as 
$\Delta G=\Delta G^o+RT\ln(\text{concentration ratio})$, where
$\Delta G^o=RT\ln(\text{equilibrium concentration ratio})$.
\end{subequations}

	Now here comes as the point of departure between the
narrative of meso-NET and the narrative of stochastic 
thermodynamics:  The former evokes Onsager's linear 
force-flux relation
\begin{equation}
   J_i(\vx,t) = -\sum_j  D_{ij}(\vx)\frac{\partial}{\partial 
             x_j} \ln\left(\frac{f(\vx,t)}{\pi(\vx)}\right).
\label{gfl}
\end{equation}
Note this relation is simply a generalized Fick's law.
Then it is well-known that combining continuity equation
(\ref{conti-eq}) with Fick's law (\ref{gfl}) yields 
the Fokker-Planck equation.

	To ``justify'' the stochastic descriptions of mesoscopic 
dynamics is one of the fundamental tasks of statistical 
physics.  Beside this meso-NET approach, there are many others: 
Bogoliubov-Born-Green-Kirkwood-Yvon (BBGKY) 
hierarchy, Mori-Zwanzig (MZ) projection, Markov partition 
and Kolmogorov-Sinai entropy method, etc.  Since the meso-NET
approach is based on Onsager's linear relation, its validity is
limited in the linear irreversible regime.

	We note that for a discrete $p_\alpha(t)$:
\begin{equation}
   \sigma^{(na)}_{\alpha\beta} =  \Big( J_{\alpha\rightarrow\beta}-
          J_{\beta\rightarrow\alpha} \Big)\Delta\mu_{\alpha\beta},
   \   \Delta\mu_{\alpha\beta} = \ln\left(\frac{p_{\alpha}
             \pi_{\beta}}{\pi_{\alpha}p_{\beta} } \right).
\end{equation}
However, the flux $J$ is \underline{not} linearly related to the 
thermodynamic potential difference $\Delta\mu$.
The same logic will not work for the stochastic dynamics of
mesoscopic chemical kinetics.

	The linear Fick's law and its alike are phenomenological
relations.  Derivation of the macroscopic diffusion equation based 
on this line of arguments naturally leads to more general
{\em nonlinear diffusion equation} \cite{goldenfeld-book,GPV}.  
The Fokker-Planck equation in phase space, however, has a much 
more fundamental origin: It is a mathematical consequence 
of Chapman-Kolmogorov's integral equation for any 
Markov process with continuous path 
in $\mathbb{R}^n$ \cite{gardiner-book}.  
In current stochastic thermodynamics, one 
takes the stochastic, Markov dynamics as given.  Nonequilibrium 
thermodynamics is not used as a justification for stochastic 
descriptions of fluctuations; rather, one attempts to derive nonequilibrium 
thermodynamics as a mathematical consequence of stochastic dynamics, 
in either continuous or discrete state space.  As a matter of fact, base 
on such a {\em Markov process hypothesis}, one can predict the
Onsager's linear force-flux relation, e.g., Fick's law, for stochastic 
dynamics in continuous space, but it also predicts a non-linear force-flux 
relation (e.g., Eq. \ref{tanh-nonl-or} below) for stochastic dynamics with 
discrete state space, such as mesoscopic chemical kinetics.

\subsection{Stochastic Liouville dynamics}
\label{sec:sld}

Bergmann and Lebowitz \cite{BL-55,LB-57}
based their new approach to nonequilibrium processes in the
phase space on a Hamiltonian mechanical system that is in 
contact with one, or multiple heat baths:
\begin{subequations}
\begin{equation}
 \frac{\partial p(\vx,t)}{\partial t} + \Big\{ p(\vx,t),H(\vx)\Big\}_{\vx}
      = \int \Big[ K(\vx,\vx')p(\vx',t)-
                    K(\vx',\vx)p(\vx,t)\Big] \rd\vx'.
\end{equation}
in which $\{p,H\}$ is the Poisson bracket, and the rhs 
represents the stochastic encounter with the heat bath(s).  By
introducing Helmholtz potental function
\begin{equation}
    F\big[p(\vx)] = \int_{\mathbb{R}^n}
              p(\vx,t)\Big( H(\vx)+\beta^{-1}\ln p(\vx,t)\Big)\rd\vx,
\end{equation}
they showed non-adiabatic entropy production
$\sigma^{(na)} = -\frac{\rd F}{\rd t}\equiv$
$\tfrac{\rd S}{\rd t}-\beta\tfrac{\rd U}{\rd t} \ge 0$ 
if the heat baths have a common temperature $\beta^{-1}$.
If not, then
\begin{equation}
      \sigma^{(na)} = \frac{\rd S}{\rd t} -\sum_{i=1}^K
              \beta_i \Phi_i,
\end{equation}
where $\Phi_i$ is the mean rate of energy flow from  
the $i$th reservoir to the system: 
$\tfrac{\rd U}{\rd t} = \sum_{i=1}^K \Phi_i$.  Furthermore,
in a nonequilibrium stationary state, 
$\sigma^{(na)} = \sum_{i,j=1}^{K'}\big(\beta_i-\beta_j\big)\Phi_j$,
where $K'<K$ are linearly independent number of energy fluxes.

\end{subequations}

	One notices that the logical development of the mesoscopic NET 
in phase space is more in line with Boltzmann's mechanical theory of 
heat than with the top-down phenomenological approach described 
by van Kampen, as discussed in Sec. \ref{sec:meso-d-ck}.

\section{Mesoscopic stochastic NET and Hill's cycle kinetics}
\label{sec:meso-stoch-net}

Most of the mathematical presentations of the mesoscopic,
stochastic nonequilibrium thermodynamics (stoch-NET)
can be carried out in either discrete or continuous state
space.  For simplicity, however, we consider a Markov dynamics 
with discrete state space $\mathscr{S}$:
\begin{equation}
  \frac{\rd p_{\alpha}(t)}{\rd t} 
   = \sum_{\beta\in\mathscr{S}} \Big(
        p_{\beta}(t)q_{\beta\alpha}-p_{\alpha}(t)
              q_{\alpha\beta}\Big).
\label{the-mastereq}
\end{equation}
We further assume that the Markov process is irreducible
and $q_{\alpha\beta}=0$ if and only if $q_{\beta\alpha}=0$.
With these assumptions, the Markov system has a unique,
positive stationary distribution $\{\pi_{\alpha}\}$ that satisfies
\begin{equation}
          \sum_{\beta\in\mathscr{S}}\Big(
            \pi_{\beta}q_{\beta\alpha}-\pi_{\alpha}
              q_{\alpha\beta}\Big) = 0,  \  \forall\alpha\in\mathscr{S},
       \  \sum_{\alpha\in\mathscr{S}}\pi_{\alpha} = 1.
\end{equation}

\subsection{Stationary distribution generates an entropic force}

	Since mathematicians can prove the existence of the
$\pi_{\alpha}$ and its positivity, one can introduce 
\begin{equation}
   E_{\alpha} = -\ln\pi_{\alpha}.
\label{p-entropic-force}
\end{equation}
This mathematical definition formalizes the notion of 
an ``entropic'' (statistical) force that does not cause the motion 
mechanically, yet it is a {\em thermodynamic force} precisely as 
articulated by Onsager.  Such a force can do mechanical work, as
has been illustrated by the polymer dynamic theory of rubber elasticity.
Of course, as any thermodynamic concept, it can have many 
different mechanistic origin.  Still, as we shall show, recognizing this 
novel ``law of force'' leads to great insights and consistency.
How to measure it, e.g., whether such an entropic force is 
onservable is an entirely different matter.  Many
researchers have discussed such nonequilibrium steady state
potential in the past, see
\cite{haken-graham,kubo,nicolis-lefevere,yin-ao,feng-wang-11}.

	With the energy given in (\ref{p-entropic-force}), one can 
introduce generalized entropy and free energy in a Markov System.
We use the term ``generalized'' to emphasize that these quantities
exist independent of whether a mesoscopic system is in an equilibrium
or not, stationary or not.  We assume the Markov dynamics has a unique stationary (invariant) distribution. This means that there is a probability based “force” pushing a system from low probability to high probability.

\subsection{Two mesoscopic laws and three nonnegative
quantities}

\begin{subequations}

	If the system has a probability distribution
$p_{\alpha}$, then the mean internal energy and 
entropy are
\begin{equation}
    \overline{E} = \sum_{\alpha\in\mathscr{S}}
              p_{\alpha}E_{\alpha}, \  
    S = -\sum_{\alpha\in\mathscr{S}}
              p_{\alpha}\ln p_{\alpha},
\end{equation}
Their difference is the Helmholtz free energy
\begin{equation}
    F\big[\{p_{\alpha}\} \big] = 
         \overline{E}(t)-S(t) = 
          \sum_{\alpha\in\mathscr{S}}
              p_{\alpha}\ln\left(
         \frac{p_{\alpha}}{\pi_{\alpha}}\right).
\end{equation}
It is easy to show that $F\ge 0$.
\end{subequations}

\begin{subequations}
	Now following the similar steps ii) to iv) in Sec. 
\ref{sec:mesonet}, assuming $p_{\alpha}(t)$ follows
the master equation in (\ref{the-mastereq}), we have
\label{dfdteq}
\begin{eqnarray}
  \frac{\rd}{\rd t} F\big[\{p_\alpha(t)\}\big]
 &=&  E_{in}\big[\{p_\alpha(t)\}\big]
           - e_p\big[\{p_\alpha(t)\}\big],
\\
   E_{in}\big[\{p_\alpha\}\big] &\equiv&
       \frac{1}{2}\sum_{\alpha,\beta\in\mathscr{S}}
       \Big(p_{\alpha}q_{\alpha\beta}-p_{\beta}
           q_{\beta\alpha}\Big)\ln\left(\frac{
             \pi_{\alpha}q_{\alpha\beta}}{\pi_{\beta}
           q_{\beta\alpha}}\right),
\\
		e_p \big[\{p_\alpha\}\big] &\equiv&
      \frac{1}{2}\sum_{\alpha,\beta\in\mathscr{S}}
    \Big(p_{\alpha}q_{\alpha\beta}-p_{\beta}
           q_{\beta\alpha}\Big)\ln\left(\frac{
             p_{\alpha}q_{\alpha\beta}}{ p_{\beta}
           q_{\beta\alpha}}\right).
\end{eqnarray}
One can in fact prove that both $E_{in}\ge 0$ and $e_p\ge 0$.
The nonnegative $E_{in}$ has also been called adiabatic entropy 
production or house-keeping heat, the nonnegative 
$e_p$ is called total entropy production rate.

Eq. \ref{dfdteq}a is interpreted  as a {\em mesoscopic free energy 
balance equation}, with $E_{in}$ being
the instantaneous energy input rate, a source term, and $e_p$ 
being the instantaneous entropy production rate, a sink.
\end{subequations}

One can also prove
\begin{equation}
    \frac{\rd F}{\rd t}\le 0.
\label{dfdtineq}
\end{equation}
The nonnegative $-\tfrac{\rd}{\rd t}F$ has also been 
called non-adiabatic entropy production or free energy dissipation.
Then Eq. \ref{dfdtineq} is like the Second Law of
Thermodynamics.  This is a very old mathematics result on 
Markov processes, which has been re-discovered many times
\cite{BL-55,moran-61,morimoto-63,voigt-81,cohen-93,qian-pre-2001}.

	Eq. \ref{dfdteq} can be re-arranged into 
$e_p = -\tfrac{\rd}{\rd t}F+E_{in}$, which offers a 
different interpretation:  The total entropy production 
$e_p$ consists of two parts: $-\tfrac{\rd}{\rd t}F\ge 0$
reflects Clausius and Boltzmann's thesis of irreversibility, and 
$E_{in}\ge 0$ reflects the existence of nonequilibrium
steady state (NESS), which is central to Nicolis and 
Prigogine's dissipative structure \cite{np-book}.

	Eqs. (\ref{dfdteq}) and (\ref{dfdtineq}) are two laws of
mesoscopic stochastic (Markov) dynamics.   They
consist of one equation (\ref{dfdteq}) and three nonnegative
quantities: $F$, $E_{in}$, $-\tfrac{\rd}{\rd t}F \ge 0$.
Then $e_p\ge 0$ follows.

\subsection{The significance of free energy balance equation}

	It is clear that the free energy balance equation
(\ref{dfdteq}) is simply an alternative expression of 
the celebrated entropy balance equation \cite{dm-book,np-book}
that lies at the foundation of classical NET:
\begin{subequations}
\label{dsdteq}
\begin{eqnarray}
  \frac{\rd}{\rd t}S\big[\{p_\alpha(t)\}\big]
 &=& e_p\big[\{p_\alpha(t)\}\big] +
         E_{ex}\big[\{p_\alpha(t)\}\big],
\\
	  E_{ex}\big[\{p_\alpha\}\big] &=& -E_{in}
     + \frac{\rd\overline{E}}{\rd t} 
\\
	&=& \frac{1}{2}\sum_{\alpha,\beta\in\mathscr{S}}
    \Big(p_{\alpha}q_{\alpha\beta}-p_{\beta}
           q_{\beta\alpha}\Big)\ln\left(\frac{q_{\beta\alpha}}{
             q_{\alpha\beta}}\right).
\end{eqnarray}
Therefore, for a system with $q_{\alpha\beta}=q_{\beta\alpha}$,
such as a microcanonical ensemble, $E_{ex}\equiv 0$ and
entropy production is the same as entropy change.  In general,
however, $E_{ex}$ does not have a definitive sign.

	For systems that are in contact with external reservoir(s),
it is well known from classical thermodynamics that entropy
is not an appropriate thermodynamics potential;  free energy
is.  For stoch-NET, the significant advantage of the free energy 
balance equation in  (\ref{dfdteq}) over the entropy balance 
equation in (\ref{dsdteq}) is obvious.   Since both $E_{in},e_p\ge 0$,
they can be definitively identified as the source and 
the sink, respectively, for the free energy.

	When a Markov system is detail balanced, e.g.,
$\pi_{\alpha}q_{\alpha\beta}=\pi_{\beta}q_{\beta\alpha}$
$\forall\alpha,\beta\in\mathscr{S}$, $E_{in}\equiv 0$
and $E_{ex}=\tfrac{\rd\overline{E}}{\rd t}$.
Such a system is like a closed system which approaches
to an equilibrium steady state with $e_p=0$.  The last
term on the rhs of the entropy balance equation
(\ref{dsdteq}a) can be moved to the lhs, and 
combined with $\tfrac{\rd S}{\rd t}$.  This yields
precisely the free energy balance equation (\ref{dfdteq}a)! 

For systems without detailed balance, a nonequilibrium 
steady state has $E_{in}=e_p\neq 0$.  They correspond
to {\em open, driven systems} with sustained transport.

	Chemists have always known that free energy
balance is different from mechanical energy conservation:
since the former involves an entropic component.  Still,
people are taught to read the ``calorie label'' on food
products in a supermarket; the validity of this practice testifies 
the significance of a rigorous {\em free energy balance 
equation}.

\end{subequations}

\subsection{Kinetic cycles and cycle kinetics}

steady state entropy production rate
\begin{eqnarray}
	e_p^{ss} &=& \sum_{i>j\in\mathscr{S}}
      \Big(J^{ss}_{ij}-J^{ss}_{ji}\Big)\ln\left(\frac{J^{ss}_{ij}}{J^{ss}_{ji}}\right)
\nonumber\\
	&=& \sum_{c:\text{ all cycles}} 
   \Big(J_c^+-J_c^-\Big)^{ss}\ln\left(\frac{J_c^+}{J_c^-}\right)^{ss}
\nonumber\\
   &=& \sum_{c:\text{ all cycles}} 
   \Big(J_c^+-J_c^-\Big)^{ss}\ln\left(\frac{q_{c_1c_2}q_{c_2c_3}\cdots
            q_{c_\kappa c_1}}{q_{c_1c_\kappa}q_{c_\kappa c_{\kappa-1}}\cdots
            q_{c_2c_1} }\right).
\label{sscycle}
\end{eqnarray}
In Eq. \ref{sscycle}, the kinetic cycle $c$ with $\kappa+1$
steps consists of the sequence of states
$\{c_1,c_2,\cdots,c_{\kappa},c_1\}$.
One should wonder what the purpose is to re-express the NESS
$e_p^{ss}$ as in (\ref{sscycle}); it contains great many more terms
since the number of cycles in a Markov graph is much more than 
the number of edges (transitions).  We note, however, that both 
$J^{ss}_{ij}=p^{ss}_iq_{ij}$ and $J^{ss}_{ji}=p^{ss}_jq_{ji}$ 
are functions of the probabilities $p^{ss}_i$ and $p^{ss}_j$.
But $(J_c^+/J_c^-)^{ss}$ is independent of any probability.
Entropy production per cycle, also called cycle affinity, is the key 
to mesoscopic, stochastic nonequilibrium thermodynamics!
The term $(J_c^+-J_c^-)^{ss}$ is simply a kinematics term:
It counts the numbers of different cycles the system passing 
through per unit time.   Recognizing this, it is immediately 
clear that one can also introduce a fluctuating entropy production
along a stochastic trajectory, counting completed cycles 
stochastically one at a time \cite{hill-chen-75}.

The cycle representation of entropy production in (\ref{sscycle}) 
was first proposed and computationally demonstrated by Hill and 
Chen \cite{hill-chen-75}.  Later, Hill's diagram approach was shown 
to be equivalent to a Markov jump process, for which 
Eq. \ref{sscycle} can be proven mathematically \cite{qqq-81,qq-82}.
A trajectory-based stochastic entropy production was 
introduced by Qian and Qian for discrete Markov processes 
as well as continuous diffusion processes in 1985 \cite{qq-85}.

\subsection{Nonlinear force-flux relation}

One of the significant results of mesoscopic, stoch-NET
is a nonlinear force-flux
relationship:  The net probability flux between states $\alpha$
and $\beta$ is $(J_{\alpha\rightarrow\beta}-J_{\beta\rightarrow\alpha})$,
and the chemical potential difference is $\Delta\mu=$
$k_BT\ln(J_{\alpha\rightarrow\beta}/J_{\beta\rightarrow\alpha})$.
Only when $\Delta\mu\ll k_BT$, e.g., 
$J_{\alpha\rightarrow\beta}\approx J_{\beta\rightarrow\alpha}$,
one has a linear relation
\begin{eqnarray}
 J_{\alpha\rightarrow\beta}-J_{\beta\rightarrow\alpha}
   &=& J_{\beta\rightarrow\alpha}\left(
        \frac{J_{\alpha\rightarrow\beta}}{
         J_{\beta\rightarrow\alpha}} -1 \right)
     \  =\   J_{\beta\rightarrow\alpha}\left(
          e^{\frac{\Delta\mu}{k_BT}}-1 \right) 
\nonumber\\
          &\approx&  J_{\beta\rightarrow\alpha} \times
            \left(\frac{\Delta\mu}{k_BT}\right).
\end{eqnarray}
In fact, an exact nonlinear force-flux relationship exists:
\begin{equation}
   J_{\alpha\rightarrow\beta}-J_{\beta\rightarrow\alpha}
  = \big(J_{\alpha\rightarrow\beta}
               +J_{\beta\rightarrow\alpha}\big)
              \tanh\left(\frac{\Delta\mu}{2k_BT}\right).
\label{tanh-nonl-or}
\end{equation}

\subsection{Comparison with macroscopic NET}

	In comparison with the macroscopic NET \cite{dm-book},
the present theory, stoc-NET, is a theory based on free energy, 
not entropy {\em per se}.   This is expected since the Markov
description represents a dynamical system with many
degrees of freedom collected under the assumption of
``random effects''.  It is a dynamic counterpart of Gibbs'
canonical ensemble rather than microcanonical
ensemble.

		While the macro-NET uses local equilibrium 
assumption to introduce the entropy function and secures
Gibbs' equation (\ref{gibbs-eq}), the stoc-NET introduces 
entropy function following Boltzmann's and Shannon's 
fundamental insights: It is a functional of the probability
distribution in the phase space.  This approach allows one
to mathematical derive a mesoscopic entropy balance equation,
and closely related free energy balance equation. Stoc-NET
does not require a local equilibrium assumption, though 
it assumes the dynamics being Markovian.  Recall that
Boltzmann's mechanical theory of heat assumes 
Newton's equation of motion \cite{gallavotti-book}.

	One of the most significant differences between stoc-NET and 
macro-NET is that one obtains self-contained mathematical
expressions for the entropy flux and entropy production rate in the 
former.   One can further mathematically prove the entropy 
production rate being nonnegative!

	Finally, but not the least, the stoc-NET has an entire
theory of fluctuating entropy production along stochastic 
trajectories, e.g., fluctuation theorems and Jarzynski-Crooks 
equalities, that provides fundamental characterizations of 
thermodynamics of small systems \cite{jarzynski,seifert,vdb-e}.

\section{Further development and applications}

\subsection{Nonequilibrium steady state and dissipative structure}

\begin{figure}[h]
\vskip 0.7cm
\begin{center}
\includegraphics[width=3in]{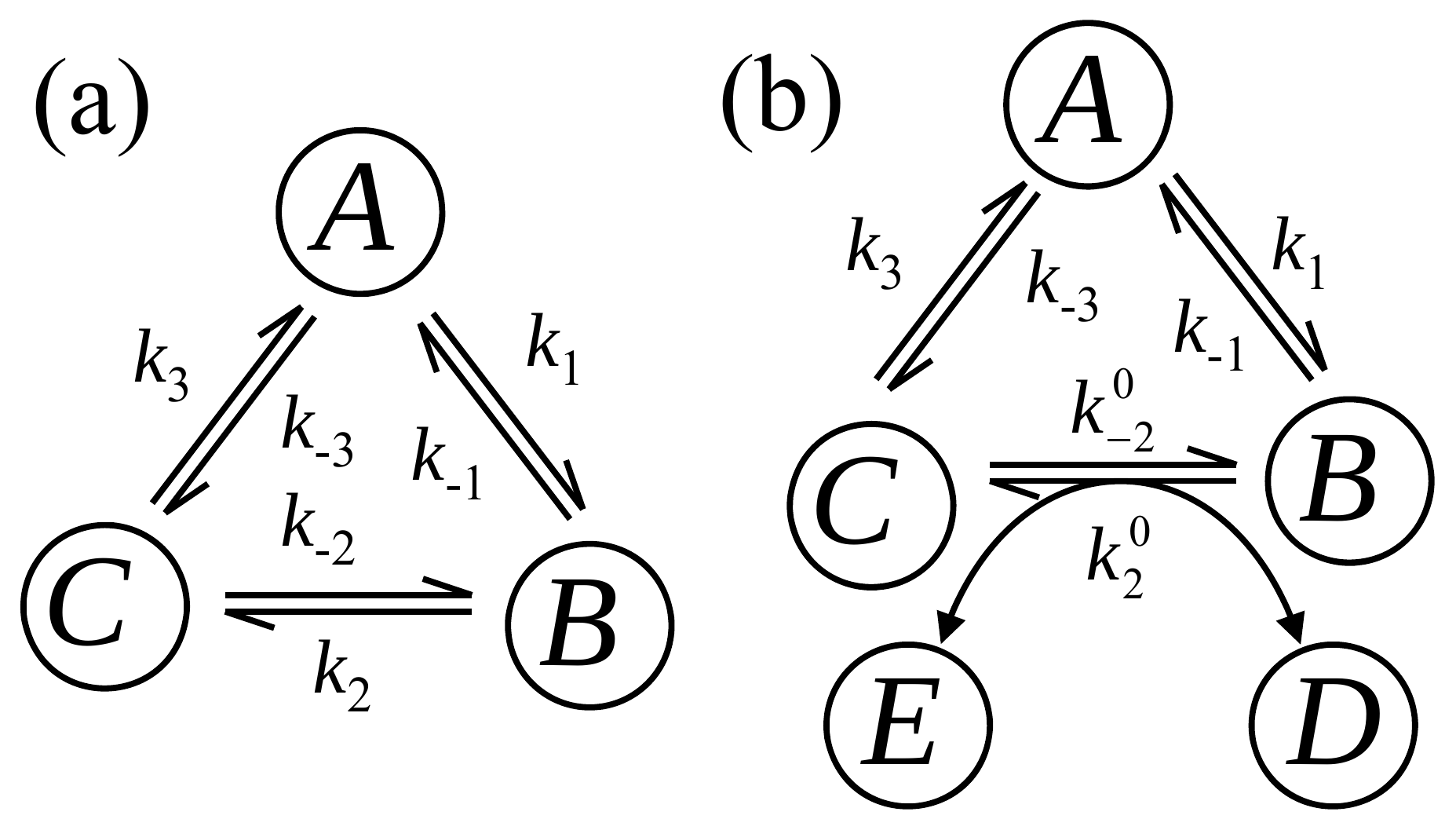}
\end{center}
\caption{(a) Unimolecular reactions and (b) pseudo-unimolecular
reactions with fixed concentrations $c_D$ and $c_E$ for species $D$ 
and $E$. One can map the nonlinear reaction on the right to the left with $k_2^oc_D=k_2$ and $k_{-2}^oc_E=k_{-2}$.  They are called 
pseudo-first-order rate constants.  If $c_D$ and $c_E$ are in their
chemical equilibrium 
$c_E/c_D=k_1k_2^ok_3/(k_{-1}k_{-2}^ok_{-3})$, then $k_1k_2k_3=k_{-1}k_{-2}k_{-3}$.
Furthermore, denoting $\gamma= k_1k_2k_3/(k_{-1}k_{-2}k_{-3})$, 
then $k_BT\ln\gamma = \mu_D-\mu_E$.  When $\gamma>1$, there is a clockwise cycle flux; and when $\gamma<1$, there is a 
counter-clockwise cycle flux.
}
\label{fig_1}
\end{figure}

Intracellular biology can be roughly thought as a complex biochemical reaction system carried out by enzymes.
The cycle kinetics of enzymes in cellular biochemistry
defines equilibrium vs. nonequilibrium steady-state (NESS) as 
illustrated in Fig. \ref{fig_1}.

\begin{figure}[h]
\vskip 0.7cm
\begin{center}
\includegraphics[width=4.in]{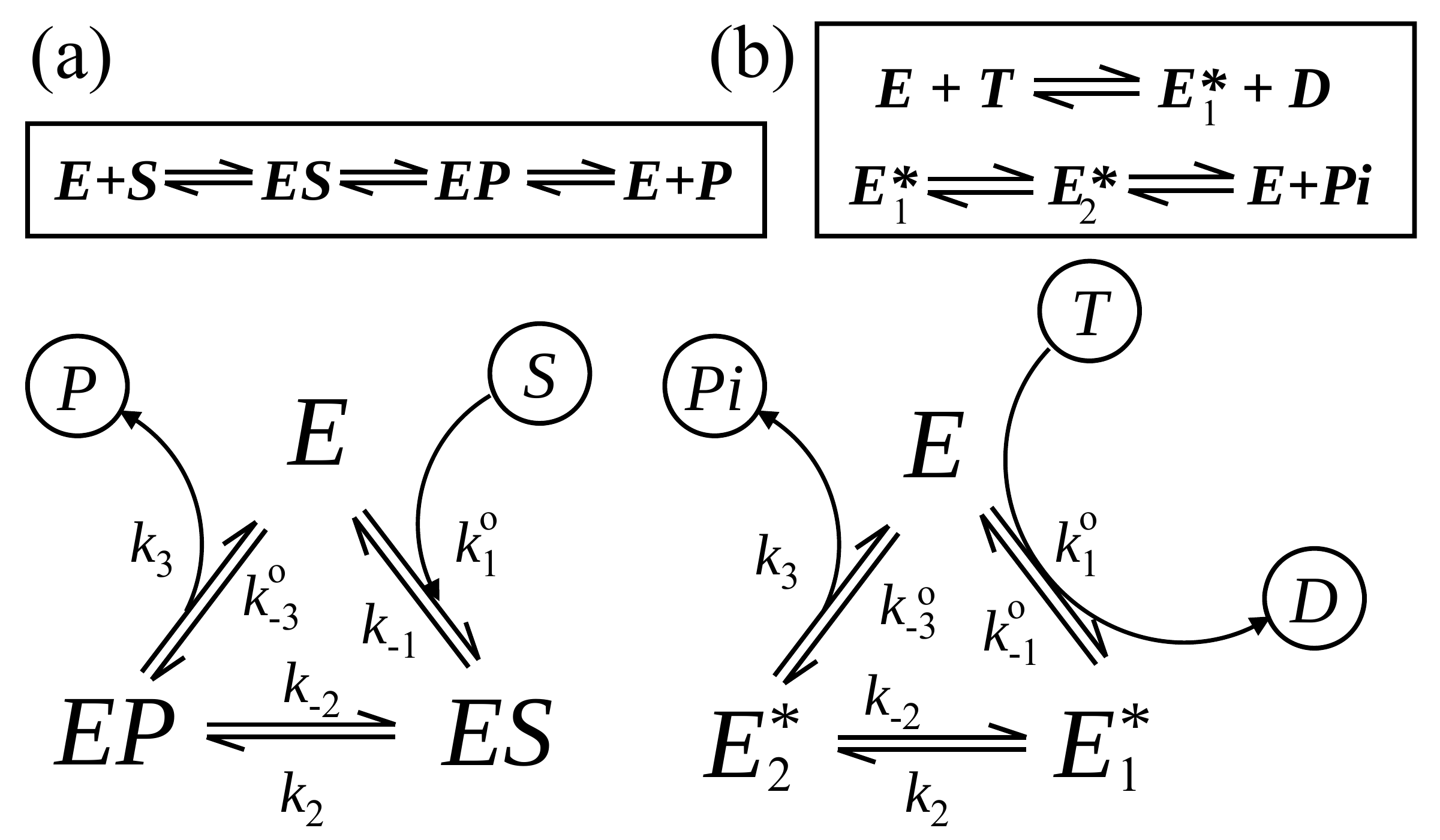}
\end{center}
\caption{(a) In cellular metabolism, almost all enzymatic
reactions are nearly irreversible.  This implies the
chemical potential of the substate $S$ is significantly
greater than that of the product $P$.  This dictates
the enzyme turnover prefers $E\rightarrow ES \rightarrow EP
\rightarrow E$ rather than other way around.  (b)  In
cellular signaling pathways, phosphorylation-dephosphorylation
cycle is one of the most widely employed biochemical mechanism
for regulating cell informations.  The cycle is driven by ATP
hydrolysis.
}
\label{fig_2}
\end{figure}

	Essentially all biochemical processes inside a living
cells are cycle kinetics.  One well-known exception is 
the creatine phosphate shuttle carried out by 
creatine kinase.  According to stoc-NET, each and 
every kinetic cycle has to be driven by a non-zero
chemical potential difference.  Fig. \ref{fig_2} are
two widely known examples.  A stationary state of an
open chemical system, sustained by a chemostatic
chemical potential difference in its environment and 
continuously dissipates free energy, epitomizes 
the notion of dissipative structure \cite{np-book}.
Rigorous mathematical theory of nonequilibrium steady 
state (NESS) and its applications can be found in
\cite{jqq,zqq,gqq}.

\subsection{Macroscopic limit: Gibbs' chemical thermodynamics}

	One of the accomplishments of Boltzmann was to obtain 
the First Law, in the form of Eq. \ref{gibbs-eq}, from Newton's
equation of motion.   But it is not often that one can 
rely on mathematics to derive a macroscopic law from 
microscopic dynamic equations, as pointed out by 
van Kampen \cite{vankampen-book}:

``Of course, the macroscopic equations cannot actually
be derived from the microscopic ones.  In practice they are
pieced together from general principles and experiences.  The
stochastic mesoscopic description must be obtained in the 
same way.  This semi-phenomenological approach is 
remarkably successful ...''

		Therefore, it is extremely satisfying when one sees
that it is possible to derive Gibbs' macroscopic, isothermal 
chemical thermodynamics from the mesoscopic 
chemical kinetic descriptions, e.g.,  Delbr\"{u}ck-Gillespie 
process \cite{ge-qian-2016-2,ge-qian-2016-3}.

\subsection{Applications: Biochemical dynamics in single cells}
\label{sec:5.3}

	Both the intracellular genetic regulatory network of self-regulating 
genes and intracellular signaling networks of phosphorylation-dephosphorylation
with substrate-activated kinase or GTPase cycle with substrate-activated 
GEF (guanine nucleotide exchange factor) are kinetic isomorphic, 
see Figure 14 of \cite{gqq} and Figure 1 in \cite{bishop-qian} for 
illustrations.  In fact, they all can be conceptually represented by the
following kinetic scheme involving autocatalysis:
\begin{equation}
       Y + \chi X  \ 
    \underset{\alpha_2}{\overset{\alpha_1}{\rightleftharpoonsfill{24pt}}}  \
   (\chi+1)X, \  \
    X \ \underset{\beta_2}{\overset{\beta_1}{\rightleftharpoonsfill{24pt}}} 
       \  Y.
\label{eq0024}
\end{equation}
For $\chi=2$ this system is closely related to the well-known
Schl\"{o}gl model in chemical kinetic literature.  The macroscopic 
kinetics follows the nonlinear differential equation
\begin{equation}
   \frac{\rd x}{\rd t} = -\frac{\rd y}{\rd t} =\alpha_1x^{\chi}y
        -\alpha_2x^{\chi+1}-\beta_1x+\beta_2y,
\end{equation}
where $x(t)$ and $y(t)$ are the concentrations of $X$ and $Y$.
Given initial values $x(0)=x_0$ and $y(0)=0$, $y(t)=x_0-x(t)$.
Non-dimensionalization of the ODE yields
\begin{equation}
\frac{\rd u}{\rd\tau} 
        = \theta u^{\chi}\big(1-u\big)- \frac{\theta u^{\chi+1}}{\mu\gamma}
             - \big(1+\mu\big) u + \mu,
\label{eq0026}
\end{equation}
where $\theta=\tfrac{\alpha_1x_0^{\chi}}{\beta_1}$,
$\mu=\tfrac{\beta_2}{\beta_1}$, and $\gamma=\tfrac{\alpha_1\beta_1}{\alpha_2\beta_2}$.  The steady state concentration of $X$,
when $\chi=0$, is \cite{qian-arpc}
\begin{equation}
     u^{ss} = \frac{\theta+\mu}{
           \theta + 1 + \mu + \theta/(\mu\gamma)},
\end{equation}
which is a monotonically increasing function of $\theta$, the
parameter representing biochemical activation.  When $\mu\ll 1$
and $\mu\gamma\gg 1$, it is nearly  $\frac{\theta}{1+\theta}$,
the expected hyperbolic curve \cite{qian-arbp}.   However, when
$\gamma=1$, $u^{ss}=\frac{\mu}{\mu+1}$ is actually independent of
$\theta$.

	For $\chi=1$, the steady-state is \cite{bishop-qian}
\begin{equation}
   u^{ss} = \frac{\theta-1-\mu+\sqrt{
            (\theta-1-\mu)^2+4\mu\theta\left[1+1/(\mu\gamma)\right]} }{2\theta\left[1+1/(\mu\gamma)\right]}.
\label{eq0028}
\end{equation}
In the very special case of $\mu=0$ and $\mu\gamma=\infty$,
$u^{ss}(\theta)$ undergoes a transcritical bifurcation at $\theta=1$.
But the transcritical bifurcation is not robust: For $\mu>0$, 
the $u^{ss}(\theta)$ in (\ref{eq0028}) is a smooth, monotonic
increasing function of $\theta$.

	For $\chi=2$, the system exhibits nonlinear bistability with saddle-node
bifurcation. When $\mu=0$ and $\mu\gamma=\infty$,
\begin{equation}
         u^{ss}_1=0 \ \textrm{ and } 
         u^{ss}_{2,3} =\frac{1\pm\sqrt{\theta^2-4\theta}}{2\theta},
\end{equation}
which exhibits saddle-node bifurcation at $\theta^*=4$. 
Saddle-node bifurcation is robust.

	For biochemical kinetics in a single cell, instead of considering
the concentration $x(t)$, one is interested in the number of $X$
molecules at time $t$, $n_X(t)$, and biochemical reactions
occur stochastically one at a time.   The Delbr\"{u}ck-Gillespie process
theory then is the appropriate mathematical representation 
of single cell biochemical kinetics and many other complex dynamics, 
just as differential equation is the appropriate mathematical 
representation of macroscopic chemical kinetics. The probability 
$p_k(t)=\Pr\big\{n_X(t)=k\big\}$ satisfies the chemical
master equation (Eq. ... in Sec. ) 
\begin{subequations}
\label{eqn0030}
\begin{equation}
   \frac{\rd}{\rd t}p_k(t) = v_{k-1}p_{k-1}(t) - 
           \big(v_k+w_k\big) p_k(t) + w_{k+1}p_{k+1}(t),
\end{equation}
in which
\begin{eqnarray}
     v_k &=& \frac{\alpha_1k(k-1)\cdots(k-\chi+1)\big(n_0-k\big)}{
            V^{\chi} } + \beta_2 \big(n_0-k\big),
\\
    w_k &=& \frac{\alpha_2k(k-1)\cdots(k-\chi)}{
            V^{\chi} } + \beta_1k,
\end{eqnarray}
where $n_0$ is the total number of $X$ and $Y$ molecules together.
\end{subequations}
The stationary probability distribution for the $n_X$ is 
\begin{equation}
  p^{ss}_k = A\prod_{\ell=1}^k \left(\frac{v_{\ell-1}}{w_{\ell}}
         \right),
\end{equation}
in which normalization factor:
\begin{equation}
    A = \left[1 + \sum_{k=1}^{n_0}\prod_{\ell=1}^k \left(\frac{v_{\ell-1}}{w_{\ell}}
         \right) \right]^{-1}.
\end{equation}
For $\chi=2$, it is easy to show that $p_k^{ss}$ has two peaks
located precisely at 

	Noting that $v_{\ell}$ and $w_{\ell}$ are both functions of the
size of a cell $V$, their macroscopic limits are 
\begin{equation}
  v(z) = \lim_{V\rightarrow\infty} \frac{v_{zV}}{V}, \ \
    w(z) = \lim_{V\rightarrow\infty} \frac{w_{zV}}{V}.
\end{equation}
In the macroscopic limit of $V,n_0\rightarrow\infty$, $\tfrac{n_0}{V}=x_0$, 
the stochastic 
process $\tfrac{n_X(t)}{V}$ becomes a smooth function of time $x(t)$,
which is the solution to the ODE $\tfrac{\rd}{\rd t} x(t) = v(x)-w(x)$.
Furthermore,

A subtle but fundamental difference: the “landscape” for a molecule is given a priori, it is the cause of the dynamics. The “landscape” for a cell is itself a consequence of cellular nonlinear, stochastic dynamics –rigorously defined via large deviation theory when 
$V\rightarrow\infty$.

\subsection{Complex systems and symmetry breaking}

\begin{figure}[h]
\vskip 0.7cm
\begin{center}
\includegraphics[width=3.2in]{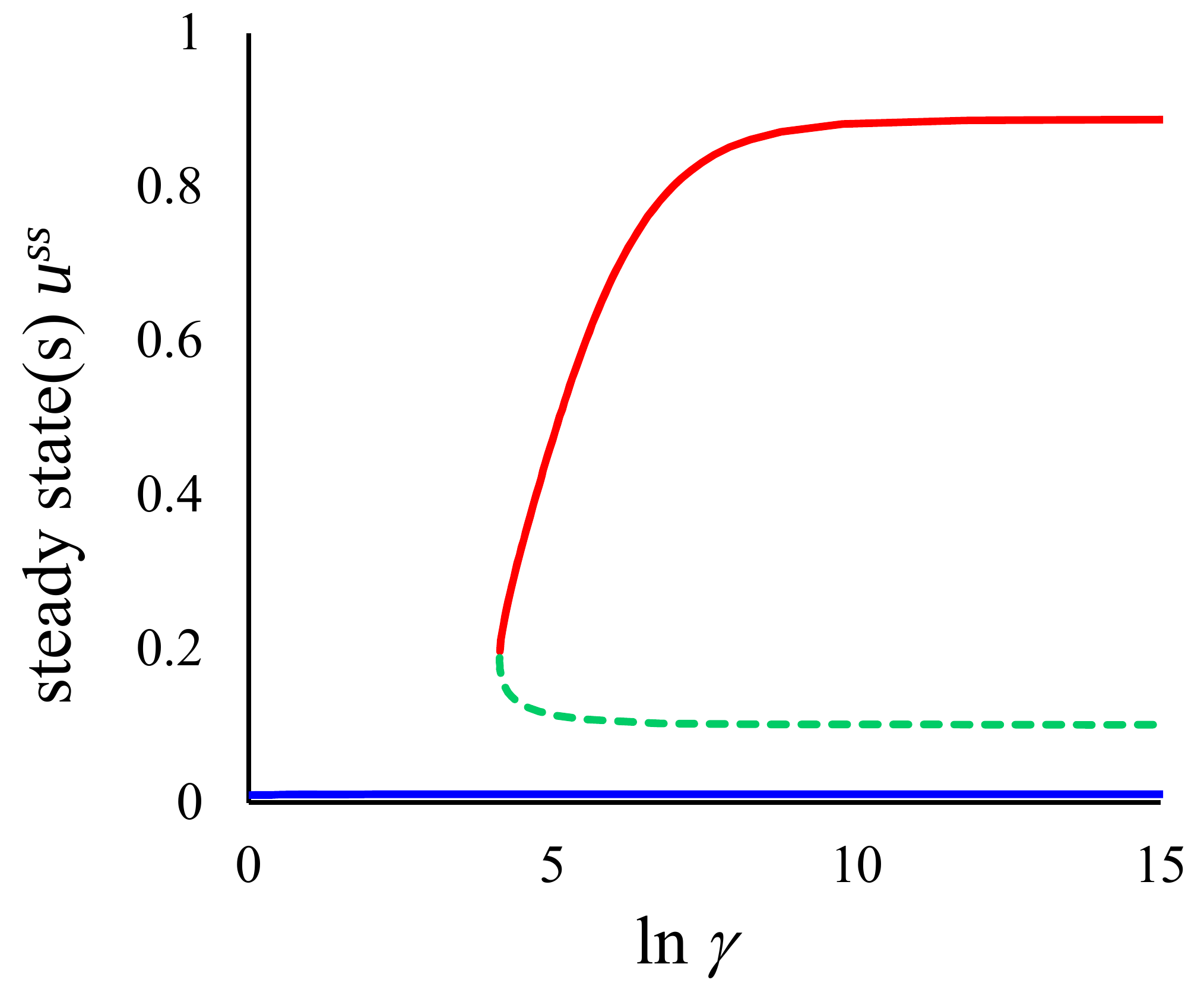}
\end{center}
\caption{Nonlinear, nonequilibrium steady state bifurcation 
gives rise to a far-from-equilibrium branch (red) in 
the ``phase diagram''.  It represents a state of the open
chemical system that is distinct from the 
``inanimate branch'' (blue) that is a  
{\em near-equilibrium continuation} of the unique chemical equilibrium 
at $\ln\gamma=0$.  The bifurcation occurs at $\ln\gamma=4.13$.
After $\ln\gamma>8$,
the ``dissipative'' red steady state is well separated
from the blue steady state by the unstable green steady state.
Steady state solutions to Eq. \ref{eq0026} are the roots of 
$\gamma=(\theta u^{\chi+1}/\mu)[\theta u^{\chi}(1-u)-(1+\mu)u+\mu]^{-1}$.  Parameter values $\mu=0.01$, 
$\theta=10$, and $\chi=2$ are used.
}
\label{fig_3}
\end{figure}

	The $u^{ss}$ given in Eq. \ref{eq0028}, which represents
the level of $X$-activity in the nonlinear, driven 
``biochemical'' kinetic system (\ref{eq0024}), is not
only a function of the level of ``activation signal'' $\theta$, but also
 a function of the amount of chemical potential difference,
$\ln\gamma$, that keeps the system away from chemical
equilibrium.  As we have already shown, when $\gamma=1$,
the $u^{ss}$ is unresponsive to $\theta$ whatsoever: ``No
dissipation, no signaling transduction'' \cite{qian-arpc}.  More
quantitatively, Fig. \ref{fig_3} shows multiple 
$u^{ss}$ as functions of the level of chemical potential 
driving force, $\ln\gamma$, for $\chi=2$.

	One could identify the red branch in Fig. \ref{fig_3} as
a ``living matter'' of the driven chemical reaction system, while
consider the blue branch, which is a continuation of the unique
equilibrium state at $\gamma=1$ with $u^{ss}$ remains near 
its equilibrium value, as an inanimate state.

	There are many schools of thoughts on complexity.  In the 
writing of condensed matter physicists \cite{pwanderson,jjh,fw_94,laughlin_wolynes}, 
{\em dynamic symmetry breaking} and {\em protected properties} or {\em rigidity}, or break-down of 
ergodicity,\footnote{Ergodicity means a 
dynamical system goes to all accessible part of the
phase space; rigidity means this motion is being
restricted to only a part of this space for a significant
amount of time. In a mathematically more careful treatment,
this happens if one takes the limit of $N\rightarrow\infty$ 
followed by $t\rightarrow\infty$.} is a key ingredient.
In most of physics of equilibrium matter, broken symmetries 
are few in number; but outside physics, with nonlinear dynamics
in an nonequilibrium systems, they are everywhere, as 
illustrated by the example in Sec. \ref{sec:5.3}.

In the case of nonequilibrium systems, the very nature of
multi-stability, and the locations of the ``attractors''
are themselves emergent phenomena, in contrast
to inert matter where the multiple attractors are dictated
by symmetry in the law of motions.
In fact, Hopfield's ``dynamic symmetry 
breaking'' can be identified in a NESS with substantial non-zero 
transport fluxes that break the detailed balance and 
time reversal symmetry.  Indeed, ``there is a sharp and accurate
analogy between the breaking-up of this [high energy world]
ultimate symmetry to give the complex spectrum of interactions
and particles we actually know and the more visible complexities''
\cite{anderson-stein}. ``At some point, we have to stop talking about
decreasing symmetry and start calling it 
increasing complication'' \cite{pwanderson}, 
or {\em complexity}. 

One of the important ideas in phase transition is
the {\em order parameter}, first proposed by L. D. Landau.
For nonlinear system with bistability, there is necessarily a 
saddle point\footnote{The term saddle has two very different meanings: it describes a point in a landscape, and it also describes a point in a vector field.  Fortunately, according to the Freidlin-Wentzell theory,
these two concepts can be unified when the noise to a 
nonlinear dynamical system is sufficiently small
\cite{fw-book,nils-berglund}, or similarly a
system is sufficiently large.} and associated with which there 
is a unstable manifold that connects the two stable states.
For such a system undergoing saddle-node bifurcation,
there is necessarily a {\em hidden} pitch-fork bifurcation
\cite{aqtw}.

	The forgoing discussion seems to be consistent with the 
general tenet of this review, and the particular result in 
Fig. \ref{fig_3}. Emergent states in dissipative systems driven 
far from equilibrium, e.g., the red steady state in Fig. \ref{fig_3}, 
indeed arise from symmetry breaking, if the latter is understood
in a broad sense as in \cite{aqtw}. 
Anderson and Stein also recognized 
autocatalysis, an idea that goes even earlier to A. M. Turing,  
as a possible mechanism for the origin of life.  But they 
were cautious and asked the poignant question 
``why should dynamic instability be the general rule
in \underline{all} dissipative systems''.

	Anderson and Stein \cite{anderson-stein} also pointed out the 
overemphasis on more complex dynamics behavior, such
as those in convection cells or vortices in turbulence, 
in complexity research.   They pointed out that these behaviors 
in generally are unstable and transitory.  Our Fig. \ref{fig_3} 
indicates that even a low dimensional fixed point, which is 
much better understood, can and should be considered as a ``dissipative structure'': On a mesoscopic scale, it has 
a time irreversible, complex temporal dynamics in its 
stochastic stationary fluctuations \cite{qian-pnas-02,qian-arbp}. 

Fluctuations favor symmetry or ergodicity, and potential energy 
or force, on the other hand, prefer special arrangements.
One of profound implication of the notion of an 
``entropic force'' is that detailed,
fast dynamics generates probability, which in turn can 
be formalized as an ``emegent force'' on a different time scale.
Therefore, the entropy force and its potential, or landscape,
are themselves emergent properties of a complex dynamics
\cite{qian-ge-mcb-12}.  This is the thermodynamic force of
Onsager \cite{onsager-31}.


\section*{Acknowledgement}
I thank Hao Ge for recent collaboration that yielded refs.
\cite{ge-qian-2016-1,ge-qian-2016-2,ge-qian-2016-3}.

\end{document}